\documentclass[twocolumn,english,aps,prb,floatfix,preprintnumbers,showpacs,amsfonts,amssymb,superscriptaddress]{revtex4}
\usepackage{graphicx}
\usepackage{dcolumn}
\newcommand{\mypath}[1]{./#1}
\newcommand{\etalter}{{\it et al.}}
\newcommand{\eg}{{\it e.\ g.}}
\newcommand{\ie}{{\it i.\ e.}}

\begin{document}

\title{Study of the One- and Two-band Models for Colossal Magnetoresistive
Manganites Using The Truncated Polynomial Expansion Method}

\author{C. \c{S}en}
\affiliation{National High Magnetic Field Laboratory and Department of Physics,
Florida State University, Tallahassee, FL 32310}

\author{G. Alvarez}
\affiliation{Computer Science \& Mathematics Division, 
Oak Ridge National Laboratory, Oak Ridge, TN 37831}

\author{Y. Motome}
\affiliation{RIKEN (The Institute of Physical and Chemical Research), Saitama 351-0198, Japan}

\author{N. Furukawa}
\affiliation{Department of Physics and Mathematics, Aoyama Gakuin University, Sagamihara 229-8558, Japan}

\author{I. A. Sergienko}
\affiliation{Condensed Matter Sciences Division, Oak Ridge National Laboratory, Oak Ridge, TN 32831}
\affiliation{Department of Physics and Astronomy, The University of Tennessee, Knoxville, TN 37996}

\author{T. Schulthess}
\affiliation{Computer Science \& Mathematics Division, 
Oak Ridge National Laboratory, Oak Ridge, TN 37831}

\author{A. Moreo}
\affiliation{Condensed Matter Sciences Division, Oak Ridge National Laboratory, Oak Ridge, TN 32831}
\affiliation{Department of Physics and Astronomy, The University of Tennessee, Knoxville, TN 37996}

\author{E. Dagotto}
\affiliation{Condensed Matter Sciences Division, Oak Ridge National Laboratory, Oak Ridge, TN 32831}
\affiliation{Department of Physics and Astronomy, The University of Tennessee, Knoxville, TN 37996}



\date{\today}

\begin{abstract}

Considerable progress has been recently made in the theoretical understanding of the
colossal magnetoresistance (CMR) effect in manganites. The existence of
inhomogeneous states has been shown to be directly related with this phenomenon,
both in theoretical studies and experimental investigations. The analysis
of simple models with two competing states and a resistor network approximation
to calculate conductances 
has confirmed that CMR effects can be theoretically reproduced using non-uniform
clustered states. However, a direct computational 
study of the CMR effect in realistic models
has been difficult since large clusters are needed for this purpose.
In this paper, the recently proposed Truncated Polynomial Expansion method (TPEM)
for spin-fermion systems is tested using the double-exchange one-band,
with finite Hund coupling $J_{\rm H}$, and two-band, 
with infinite $J_{\rm H}$, models. 
Two dimensional lattices as large as 48$\times$48 are studied, 
far larger than those that can be handled
with standard exact diagonalization (DIAG) techniques for the fermionic sector. The
clean limit (i.e. without quenched disorder) is here analyzed in detail. Phase diagrams
are obtained, showing first-order transitions separating ferromagnetic
metallic from insulating states. A huge magnetoresistance is found at low
temperatures by including small magnetic fields, in excellent agreement with 
experiments. However, at temperatures above the Curie transition the effect is much
smaller confirming that the standard finite-temperature 
CMR phenomenon cannot be understood using homogeneous states.
By comparing results between the two methods, TPEM and DIAG, 
on small lattices, and by analyzing the systematic
behavior with increasing cluster sizes, it is concluded that the TPEM 
is accurate to handle realistic manganite models on large systems. 
Our results pave the way to a frontal computational attack 
of the colossal magnetoresistance phenomenon using double-exchange like models,
on large clusters, and including quenched disorder.

\end{abstract}

\maketitle

\section{Introduction}

The study of transition metal oxides (TMOs) has been among the most important
areas of investigations in condensed matter physics in the last two decades.
The excitement in this context started with the high-temperature superconductors
and was later followed by the discovery of the colossal magnetoresistance in
manganites,\cite{tokura,review} 
as well as a plethora of equally interesting phenomena in several
other oxides. Strong correlations ($i.e.$  large 
electron-electron  or electron-phonon couplings, or both) play a major role
in the physics of these materials. In addition, the presence of nearly degenerate states
renders some of these oxides highly susceptible to external perturbations.
In fact, TMOs appear to share a phenomenology similar to that of complex
systems, with nonlinearities and sensitivity to details.\cite{dagotto-science}

We focus in this work on the manganites, area where in recent
years considerable progress has been made, both in theory and
experiments.\cite{review} In the late 90's, it was predicted that many Mn-oxides should
be inhomogeneous at the nanoscale, due to the unveiling of tendencies toward
electronic phase separation.\cite{yunoki}  On the experimental
front, the evidence for inhomogeneous states was rapidly gathered 
and it is by now widely accepted, with building blocks
typically having small nanoscale sizes.\cite{cheong} 
Subsequent theoretical work showed that similar tendencies can also occur
after the inclusion of quenched disorder effects -- caused for instance by
chemical doping-- near first-order phase transitions.\cite{giant} The key influence
of quenched disorder was also observed in simulations of the one-band model for manganites
including cooperative phonons \cite{motome,sen} and also for two bands 
with Jahn-Teller phonons.\cite{aliaga} 
This key role of quenched disorder 
postulated by theory was observed experimentally
using a Mn-oxide compound that can 
be prepared both in crystal ordered and disordered
forms.\cite{tokura-disorder,ueda} Remarkably, only the latter presents a CMR effect.

While the presence of quenched disorder was theoretically found to generate metal insulator
transitions similar to those found experimentally,
the actual existence of large magnetotransport effects is difficult to test
in unbiased theoretical studies. Using toy models that have phase competition
and first-order transitions, and supplementing the investigations with 
a random-resistor network approximation, huge magnetoresistances were obtained
in resistance vs. temperature profiles in excellent agreement 
with experiments.\cite{burgy}
However, it is certainly desirable to carry out similar investigations in more
realistic models, of the double-exchange variety, and with quantum mechanical 
procedures
to calculate the conductances. Alas, this task is  tremendously difficult
with standard computational methods that rely on the exact diagonalization
of the fermionic sector and the Monte Carlo simulation of the classical
$t_{\rm 2g}$ spins.\cite{review} The effort in this context grows like $N^4$, with $N$
the number of sites, severely limiting the clusters that can be studied.
Since theory suggests 
that physics related with percolative phenomena is anticipated upon the
introduction of magnetic fields in nano-clustered states,\cite{review,sen} large clusters
must be used for accurate simulations. Fortunately, important progress has been made
in recent years toward the development of a novel method to carry out these
investigations.\cite{motome99,furukawa01,furukawa03,motome-disorder} The technique, TPEM, 
has a CPU time that scales like $N$
and, as a consequence, it is a promising technique for these investigations.
Previous studies have shown that the $J_{\rm H}$=$\infty$ one-band model with
and without quenched disorder can be accurately studied.\cite{motome99,furukawa01,furukawa03,motome-disorder,alvarez}
However, the method
has not been tested yet under some of the more severe circumstances needed for
a realistic study, namely the use of two active orbitals per site ($i.e.$ the
doubly degenerate $e_{\rm g}$ sector of Mn ions) and/or with a finite Hund coupling. 

It is important to remark that there are at least two other independent
methods that have been proposed to improve on
the exact diagonalization of the fermionic sector approach: (1) The hybrid
Monte Carlo technique
of Alonso {\it et al.},\cite{alonso1} inspired in lattice gauge theory,
that comfortably allows calculations on lattices up to $10^3$ sites
in the limit of $J_{\rm H}$=$\infty$ and 
for one orbital, \cite{alonso2} and with a linear with $N$ increase in effort,
and (2) the method of Kumar and Majumdar \cite{kumar} that has been already
applied to a variety of models reaching 32$\times$32 clusters for one orbital 
and at $J_{\rm H}$=$\infty$. The scaling of this method is $N^3$. 
Our choice of the TPEM is motivated by the perception that a linear
with $N$ cost is needed to handle the anticipated percolative physics
that emerges when quenched disorder and phase competition occurs. Also it
seems easier to implement than method (1) where auxiliary fields are needed. 
Nevertheless, this is
not a critique on methods (1) and (2): they should be strongly pursued as well.
Only future work can decide which method is the best for the type
of problems described here.


It is the main purpose of this paper to present a systematic study of
the TPEM applied to models that are widely believed to be realistic
for manganite investigations, in the clean limit. The conclusions indicate that the technique works
properly for these investigations, opening an avenue of research toward
the ultimate goal of conducting a fully realistic simulation of the two-band double-exchange model
including quenched disorder. The present results include a detailed analysis
of both metallic and insulating phases on lattices as large as 48$\times$48
sites, about 20 times larger in number of sites than it is  possible to handle 
with exact diagonalization techniques.
In addition, here it is discussed the existence of a huge magnetoresistance at low
temperatures, in agreement with experiments for 
$\rm(Nd_{\it 1-y} Sm_{\it y})_{1/2} Sr_{1/2} Mn O_3$.\cite{low-T-CMR}
This phenomenon was theoretically
studied before by Aliaga {\it et al.} \cite{aliaga} as well, 
although on much smaller systems. Nevertheless,
the conclusions are similar: this somewhat exotic ``low temperature'' 
CMR phenomenon can be
understood as a natural consequence of the existence  
of a first-order metal-insulator transition in the phase diagram and,
thus, a clean-limit
simulation is sufficient for this purpose.
However, in our investigations
it is also confirmed that these clean limit simulations are $not$ able to generate the
standard CMR effect above the Curie temperature using states that are homogeneous.
Quenched disorder or strain fields are likely important for this purpose, which will
be the subject of near future efforts.

The organization of the paper is simple. The theoretical aspects of the TPEM are briefly reviewed in Section~\ref{sec:tpem}.
The systematics
of the TPEM in the case of the one-band model, with emphasis on the dependence
with parameters and size effects is discussed in Section~\ref{sec:oneband}. We also present physical results related
with large magnetoresistance effects found at low temperatures, in large clusters. 
Then, in Section~\ref{sec:twoband}, the emphasis shifts to the two bands model, with an analogous study of TPEM
performance and effects of magnetic fields. Conductances are evaluated
for both models using the TPEM  and reasonable results are observed. Overall, it is
concluded that the TPEM is adequate for a frontal future attack of the CMR
phenomenon using realistic models and quenched disorder.

\section{Review of the TPEM}\label{sec:tpem}

For completeness, here a brief review of the TPEM is presented
following  closely Ref.~\onlinecite{alvarez}.
Consider a model defined by a general Hamiltonian,
$\hat{H}=\sum_{ij\alpha\beta}c^\dagger_{i\alpha} H_{i\alpha,j\beta}(\phi)c_{j\beta}$, 
where the indices $i$ and $j$ 
denote a spatial index, 
while $\alpha$ and $\beta$ are internal degree(s) of freedom, \eg\ 
spin and/or orbital. As in the case of spin-fermion models,
the Hamiltonian matrix depends on the configuration of one or more classical fields,
represented 
by $\phi$. Although no explicit indices will be used, the field(s) $\phi$ 
are supposed to have a spatial dependence. 
The partition function for this Hamiltonian is schematically given by:
$
Z=\int d\phi \sum_n \langle n | \exp(-\beta\hat{H}(\phi)+\beta\mu\hat{N}) |n\rangle
$
where $|n\rangle$ are the eigenvectors of the one-electron sector, and the $\phi$
integral denotes the integration over all the classical fields. 
Here $\beta$=$1/(k_{\rm B} T)$ is the inverse temperature.
The number of particles  (operator $\hat{N}$) is adjusted
via the chemical potential $\mu$. 

The procedure to calculate observables (energy, density, action, etc) is the following. 
First, an arbitrary configuration of classical fields $\phi$ is selected as a starting point.
The Boltzmann weight or action of that configuration, $S(\phi)$, is calculated by diagonalizing the one-electron sector. 
Then, a small local change of the field configuration is proposed, so that the new configuration
is denoted by $\phi'$ and its action is recalculated to obtain the difference in action $\Delta S = S(\phi')-S(\phi)$. 
Finally, this new configuration is accepted or rejected based on a
Monte Carlo algorithm, such as Metropolis or heat bath, and the cycle starts again. 
In summary, in the standard algorithm (that we will denote here as DIAG)
the observables are calculated using an exact 
diagonalization of the one-electron sector 
for every classical field configuration, and Monte Carlo integration for those
classical fields.\cite{review} 

The TPEM replaces the diagonalization for a
polynomial expansion as discussed below (details can be found in
Refs.~\onlinecite{motome99,furukawa01,furukawa03}). 
It will be assumed that the Hamiltonian $H(\phi)$ is ``normalized'', 
which simply implies a rescaling
in such a way that the normalized Hamiltonian has eigenvalues $\epsilon_v\in[-1,1]$.
Simple observables can be divided in two categories:
(i) those that do not depend directly on fermionic operators, \eg\ 
 the magnetization, susceptibility and classical spin-spin correlations 
 in the double
exchange model, and
(ii) those for which a function $F(x)$ exists
such that they can be written as:
$A(\phi)=\int_{-1}^{1}F(x)D(\phi,x)dx$,
where $D(\phi,\epsilon)=\sum_\nu\delta(\epsilon(\phi)-\epsilon_\nu)$, 
and $\epsilon_\nu$ are the eigenvalues of $H(\phi)$, \ie\ $D(\phi,x)$ is the density of
states of the system. 

For category (i), the calculation is 
straightforward and simply involves the average over Monte Carlo configurations.
Category (ii) includes the effective action or generalized
Boltzmann weight and this quantity is particularly important because 
it is calculated at every MC step 
to integrate the classical fields. Therefore, we first briefly review how to deal with this type of observables.
As Furukawa \etalter\ showed, it is convenient to expand the function $F(x)$ in terms of Chebyshev
polynomials in the following way:
$F(x)=\sum_{m=0}^{\infty}f_mT_m(x)$,
where $T_m(x)$ is the $m-$th Chebyshev polynomial evaluated at $x$.
Let $\alpha_m=2-\delta_{m,0}$. The coefficients $f_m$ are calculated with the formula:
$f_m=\int_{-1}^1 \alpha_m F(x)T_m(x)/(\pi\sqrt{1-x^2})$.
The moments of the density of states are defined by:
\begin{equation}
\mu_m(\phi)\equiv\sum_{\nu=1}^{N_{dim}}\langle \nu| T_m(H(\phi))|\nu\rangle,
\label{eq:moments}
\end{equation}
where $N_{dim}$ is the dimension of the one-electron sector.
Then, the observable corresponding to the function $F(x)$, can be calculated using
\begin{equation}
A(\phi)=\sum_m f_m \mu_m(\phi).
\label{eq:expansion}
\end{equation}
In practice, the sum in Eq.~(\ref{eq:expansion}) is performed up 
to a certain cutoff value
$M$, without appreciable loss in 
accuracy as described in Refs.~\onlinecite{motome99} and \onlinecite{furukawa01},
and as extensively tested in the main sections of this paper for realistic manganite
Hamiltonians.
The calculation of $\mu_m$ is carried out recursively using
$|\nu;m\rangle  =  T_m(H)|\nu\rangle=2H|\nu;m-1\rangle-|\nu;m-2\rangle$.
These same vectors are used to calculate the moments.
The process involves a sparse matrix-vector product, \eg\ in $T_m(H)|\nu\rangle$,
  yielding a cost  of
$O(N^2)$ for each configuration, \ie\ $O(N^3)$ for each Monte Carlo step. This
represents an improvement in a factor $N$ compared with DIAG.

In addition,
an extra improvement of the method described thus far 
has been proposed \cite{furukawa03} based on 
two truncations: (i) of the sparse matrix-vector product and (ii) of the difference in action for local Monte Carlo updates.
The resulting algorithm has an expected CPU time growing like $N$. 
The first truncation is possible because of the form of the 
vectors $|\nu;m\rangle$. Indeed, for $m$=$0$ these are simply basis
vectors that can be chosen with only one non-zero component. The $m$=$1$ 
vector is obtained by applying $T_1(H)$=$H$ to the basis
vector. Since $H$ is sparse (consider for instance
a nearest-neighbor hopping), the vector $|\nu;m=1\rangle$ will have
non-zero elements only in the vicinity of $\nu$. For general $m$, 
the non-zero elements will propagate in what resembles a
diffusion process. Note that we only have to keep track 
of the non-zero indices to perform the sparse matrix-vector product. Since we
are only discarding null terms, this truncation does not introduce any approximation. It is possible to go a step further and
discard not only zero elements but elements smaller than a certain 
threshold denoted by $\epsilon_{\rm pr}$. In this case, the results are
approximate, 
but the exact results are recovered in the limit $\epsilon_{\rm pr}\rightarrow0$.
In this paper, the dependence of results with various values for this cutoff
(and the one described below) is discussed.

The second truncation involves the difference in effective action,
which is calculated very frequently in the Monte Carlo integration procedure.
The function corresponding to the effective action 
for a configuration $\phi$ is defined by
$F^S(x) = -\log(1+\exp(-\beta(x-\mu)))$ and $S(\phi)$ admits an expansion as in Eq.~(\ref{eq:expansion}), with
coefficients $f^S_m$ corresponding to $F^S(x)$.
In practice, only the difference in action,
$\Delta S $=$ S(\phi')-S(\phi)$  has to be computed for
every change of classical fields. 
Since this operation requires calculating two sets of moments, for
$\phi$ and $\phi'$, the authors of Ref.~\onlinecite{furukawa03} 
have also developed a truncation procedure for this
trace operation controlled by a small parameter, $\epsilon_{\rm tr}$.
This truncation is based on the observation that if $\phi$ and 
$\phi'$ differ only in very few sites 
(as is the case with local Monte Carlo updates), then the
corresponding moments will differ only for certain 
indices allowing for a significant reduction
of the computational effort. 
Again, the exact results are recovered for $\epsilon_{\rm tr}\rightarrow0$, so this approximation
is controllable.

Another key advantage of the TPEM is that it can be parallelized,
 because 
 the calculation of the moments in Eq.~(\ref{eq:moments}) for each basis
ket $|\nu\rangle$ is independent. Thus, the basis can be partitioned in such a way that each processor calculates the moments corresponding to a portion of the basis. 
It is important to remark that the data to be moved between
different nodes are small compared to calculations in each node:
communication among nodes is mainly done here to add up all the
moments.
The possibility of parallelizing this algorithm can be contrasted with the conventional 
method where a matrix diagonalization
is performed at every Monte Carlo step; in that case the calculations must be serial because it is
difficult to make an efficient parallelization of the matrix diagonalization.

\section{RESULTS FOR THE ONE BAND MODEL}\label{sec:oneband}

\subsection{Definition}
After introducing the computational method, we will now focus on its performance
starting with the one-band model for manganites. Historically, this model was
among the first proposed for Mn-oxides and it is still widely used, although
it does not incorporate the two orbitals $e_{\rm g}$ of relevance in Mn ions.
This more involved two-band version will be studied in the next section.

The one band Hamiltonian used in this study is given by:
\begin{eqnarray}
H_{1b}&=&-t\sum_{\langle ij \rangle , \alpha} (c_{i,\alpha}^{\dag}c_{j,\alpha} + \mbox{h.c.})
-J_{\rm H}\sum_{i,\alpha, \beta}c_{i,\alpha}^{\dag} \sigma_{\alpha,\beta}c_{i,\beta} \nonumber
\\ &+&J_{\rm AF}\sum_{\langle ij \rangle}\mathbf{S}_{i}\cdot\mathbf{S}_{j},
\end{eqnarray}
where $c_{i,\alpha}^{\dag}$ creates an electron at site $i$ with spin $\alpha$,
$\mathbf{\sigma}$ are the Pauli spin matrices, $\mathbf{S}_i$ is the total spin of the
$t_{\rm 2g}$ electrons (assumed to be localized and classical), $\langle ij
\rangle$ indicates summing over nearest neighbor sites, $t$ is the nearest neighbor 
hopping amplitude for the movement of 
electrons ($t$ also sets the energy unit), 
$J_{\rm H}>0$ is the Hund coupling, and $J_{\rm AF}>0$
is the antiferromagnetic coupling between the localized spins.  The study carried out
in this manuscript is based on the $clean$ limit, namely the couplings that appear
in the Hamiltonian do not have a site index, which would be necessary if quenched
disorder is incorporated. The study of realistic models including disorder
will be carried out in a future investigation, since it represents an order of
magnitude extra effort due to the average over disorder configurations.

Here and in 
the rest of the paper, spatial indices will always be denoted without arrows or
bold letters independently of the dimension. Also the notation $i+j$ is meant to
represent the lattice site given by the vectorial sum of the vectors
corresponding to $i$ and $j$, respectively.

\subsection{TPEM performance}

\subsubsection{Test of the TPEM in Small Systems}

\begin{figure}
\centerline{
\includegraphics[clip,width=9cm]{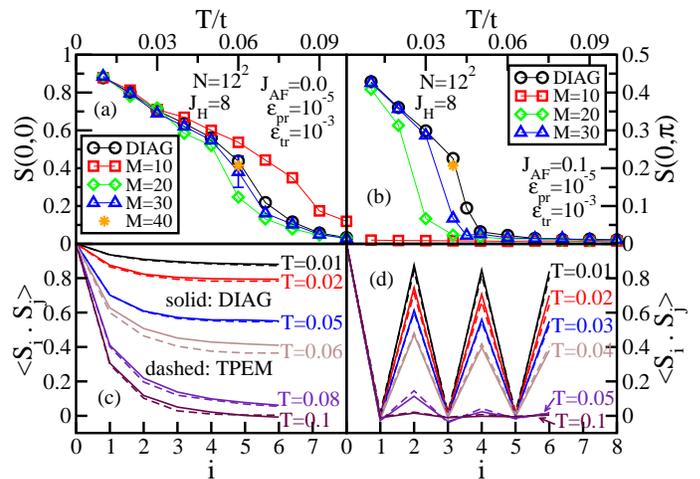}}
\caption{(color online). Dependence of the TPEM algorithm results on the number of terms M in the expansion. Shown are the spin structure
factors at the momenta characteristic of (a) the FM state and (b) the Flux state,
normalized to 1 and 0.5, respectively, for the perfect states. Results are obtained on a $12\times12$
lattice and they are compared with the numbers gathered using the exact diagonalization algorithm, all calculated at
density $\langle n\rangle=0.5$. (a),(c) correspond to $J_{\rm AF}=0.0$ and (b),(d) to $J_{\rm AF}=0.1$. Measurements were taken 
every $10$ steps of a MC run of $2000$ total iterations, after discarding $2000$ steps for thermalization. A random 
starting configuration is used for each $T$. In (a), at the most difficult temperature, $T$=0.06,
where critical fluctuations are strong, the result shown was confirmed using several
different starting configurations, including ordered ones. 
The average $S(0,0)$ obtained by this procedure was very
similar among the several starts. The shown error bars at this temperature and M=30 
mainly arise from the expected critical fluctuations.
In (b), a good convergence at $T$=0.04 is only achieved by using 40
moments with $\varepsilon_{\mbox{\scriptsize{{pr}}}}=10^{-7}$ and $\varepsilon_{\mbox{\scriptsize{{tr}}}}=10^{-5}$ 
and the result is shown with an orange star just below the exact result. In (c), the TPEM parameters used are $M=30$,
$\varepsilon_{\mbox{\scriptsize{{pr}}}}=10^{-5}$ and $\varepsilon_{\mbox{\scriptsize{{tr}}}}=10^{-3}$, except at $T$=0.06, where the
convergence is achieved by using 40 moments with $\varepsilon_{\mbox{\scriptsize{{pr}}}}=10^{-7}$ and 
$\varepsilon_{\mbox{\scriptsize{{tr}}}}=10^{-5}$. In (d), all the results shown were obtained with $M=30$,
$\varepsilon_{\mbox{\scriptsize{{pr}}}}=10^{-5}$ and $\varepsilon_{\mbox{\scriptsize{{tr}}}}=10^{-3}$, and solid lines
represent the DIAG results while the dashed lines are the TPEM results.} 
\label{Figure1}
\end{figure}

As explained before, 
the TPEM has three controlling parameters: $M$, $\epsilon_{\rm pr}$, 
and $\epsilon_{\rm tr}$.
In the limit when the first parameter runs to infinity, and the other two to zero, the exact results are
recovered. For the TPEM to be useful, accurate results must be obtained for
values for these parameters that allow for a realistic computational study. 
In Fig.~\ref{Figure1}(a), the dependence of the zero-momentum
spin structure factor, of relevance for ferromagnetism, is shown vs. temperature, using
a 12$^2$ cluster and the values of $J_{\rm H}$ and $J_{\rm AF}$ indicated. In this case,
it is expected that a FM state will form at low temperatures, as observed numerically.
Results for many values of $M$ are shown, at fixed values of $\epsilon_{\rm pr}$, and $\epsilon_{\rm tr}$. 
Clearly, $M$=10 only captures the low and high
temperature limits, but it is not accurate near the critical temperature. The results for $M$=20
are much better, but still there is a visible discrepancy near the region where $S(0,0)$ changes
the fastest. However, for $M$=30 and 40, fairly accurate results are obtained. 
In Fig.~\ref{Figure1}(c), it is shown that even the spin correlations at the largest distances
are accurately reproduced with $M$=40 terms in the expansion.

In Fig.~\ref{Figure1}(b), similar results are presented but now for the spin-structure-factor
corresponding to the ``Flux'' phase --  nearest-neighbor spins at 90 degrees
forming a staggered arrangement of nonzero plaquette fluxes.
This phase appears at $n$=0.5 with increasing $J_{\rm AF}$,
as reported in previous investigations. \cite{flux-phase} In this case, $M$=10 and even 20  
produces results dramatically different from those of DIAG. However, 
for $M$=30 discrepancies are observed
only in a range of temperatures near the critical transition, with low and high temperatures
under control. Finally, $M$=40 leads to very accurate results, as in case (a). Even the
spin correlations are under well control for this number of terms in the expansion 
(see Fig.~\ref{Figure1}(d)). The range $M$$\sim$30-40 appears systematically in our
investigations, and it is expected to provide safe values of $M$ for studies  
of the type of
spin-fermion models under investigation in manganites.

\begin{figure}
\centerline{
\includegraphics[clip,width=9cm]{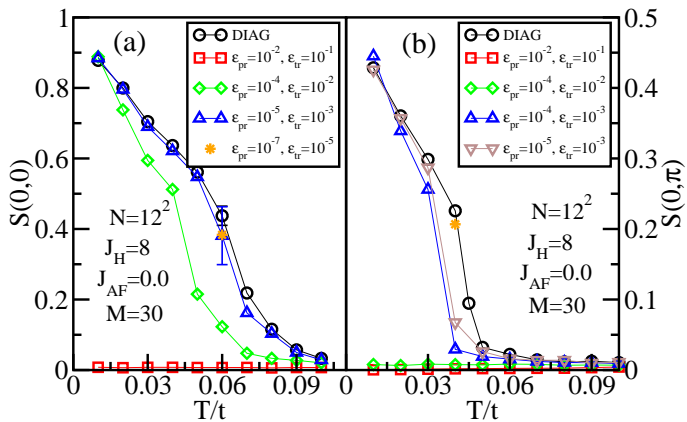}}
\caption{(color online). $\varepsilon$ dependence of the TPEM algorithm  results (M=30) 
for a $12\times12$ lattice, compared with the
exact (DIAG) results. Both  are obtained  at density $\langle n \rangle$=0.5, using (a) $J_{\rm AF}=0.0$ 
and (b) $J_{\rm AF}=0.1$. Measurements were taken at every $10$ steps of 
a $2000$ MC steps total run, 
after $2000$ steps for thermalization. In (b), the convergence at $T$=0.04 is achieved by using 40
moments with $\varepsilon_{\mbox{\scriptsize{{pr}}}}=10^{-7}$ and $\varepsilon_{\mbox{\scriptsize{{tr}}}}=10^{-5}$, 
as shown with an orange star just below the exact result. 
The starting configurations and error bar convention is as in Fig.\ref{Figure1}.} 
\label{Figure2}
\end{figure}

In Fig.~\ref{Figure2}, a study of the dependence 
of results with $\epsilon_{\rm pr}$ and $\epsilon_{\rm tr}$ is
presented, working at $M$=30. From Fig.~\ref{Figure2},
clearly there are large $\epsilon$'s that lead to unphysical results, but with 
decreasing values an accurate evaluation of observables is reached. 
In this and other investigations, values such as $\epsilon_{\rm pr}$=10$^{-5}$ 
and $\epsilon_{\rm tr}$=10$^{-3}$ are generally found to be accurate, 
with only a few exceptions.

\subsubsection{Dependence of Results on Lattice Sizes}

\begin{figure}
\centerline{
\includegraphics[clip,width=9cm]{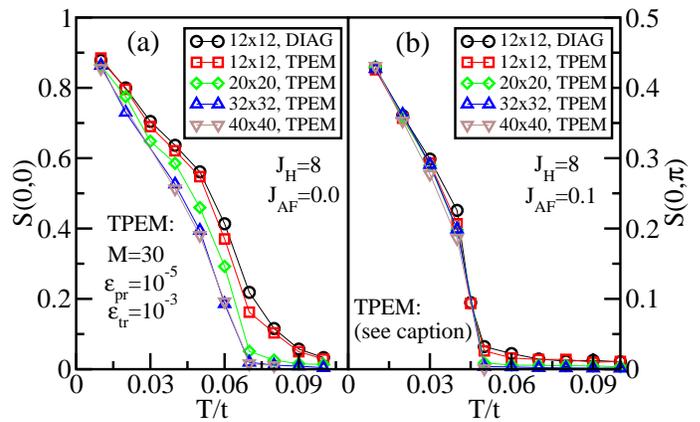}}
\caption{(color online). Lattice size dependence of the TPEM algorithm results 
for the lattices and parameters shown, working at density $\langle n \rangle$=0.5 and using
(a) $J_{\rm AF}=0.0$ and (b) $J_{\rm AF}=0.1$. 
Measurements were taken at every $10$ steps of a
$2000$ MC total steps run, after discarding $2000$ MC steps for thermalization. 
For the 20$\times$20 lattice and larger, the starting configuration
used was a perfect FM state.
In (b), the starting configuration used is a random one except at $T$=0.04 where the starting configuration is chosen to be a perfect flux
state for faster convergence. In addition, for lattices except $40\times40$, the convergence at $T$=0.04 is achieved by using 40
moments with $\varepsilon_{\mbox{\scriptsize{{pr}}}}=10^{-7}$ and $\varepsilon_{\mbox{\scriptsize{{tr}}}}=10^{-5}$, while for other temperatures 
$M=30$ with $\varepsilon_{\mbox{\scriptsize{{pr}}}}=10^{-5}$ and $\varepsilon_{\mbox{\scriptsize{{tr}}}}=10^{-3}$ were sufficient. For the lattice 
$40\times40$, $M=40$ with $\varepsilon_{\mbox{\scriptsize{{pr}}}}=10^{-6}$ and $\varepsilon_{\mbox{\scriptsize{{tr}}}}=10^{-4}$ were used for all
temperatures.} 
\label{Figure3}
\end{figure}

An approximate method that depends on some parameters, such as in the case of the TPEM, is
practical only if by fixing those parameters on small systems, their values still provide
accurate numbers as the lattice sizes increase. A qualitative way to carry out this test
is to perform the studies on large clusters and see that all the trends and approximate
numbers remain close to those known to be accurately 
obtained on small systems, or expected from other techniques or
physical argumentations. Figure \ref{Figure3}(a,b)
supports the notion that TPEM indeed behaves properly in this respect, namely the
range of $M$ and $\epsilon$'s identified in the previous subsection are sufficient to
produce qualitatively similar results even when the number of sites grows by a factor 10. 
In (a), the expected size dependence corresponding to a second order FM transition is
found. For a 40$\times$40 cluster, the Curie temperature appears located
at $T$$\sim$0.07. In (b), the size dependence is almost negligible. The transition is 
far sharper for the paramagnetic-flux transition, as already noticed in Fig.~\ref{Figure2}(b). This is an intriguing feature that will be investigated in future work:
while the first-order low-temperature metal-insulator transitions are clear and
well established in realistic models for manganites, the presence of first-order
transitions between ordered and disordered phases varying temperature is far less
obvious, and TPEM studies on large lattices can properly address this
issue. For our current purposes, here it is sufficient to state that
the TPEM appears to behave properly with increasing lattice sizes, both in
metallic and insulating regimes.

At this point a clarification is important. In principle, two-dimensional systems should not show
true critical temperatures due to the Mermin-Wagner theorem. 
However, it is well known that
in systems where the Mermin-Wagner theorem applies, such as the two-dimensional NN Heisenberg model,
the antiferromagnetic spin correlations exponentially diverge with decreasing temperature.
An exponential behavior, defines via the exponent 
a temperature scale $T^*$ below which the correlations are much larger than
any lattice size that can be practically studied numerically. This may seem like a problem,
but it is not: very large correlation lengths also render the system {\it very susceptible to small
perturbations}. In particular, we have shown that tiny deviations from the fully symmetric
Heisenberg model, such as introducing Ising anisotropies, stabilize $T^*$ into a true
critical temperature. In fact, simulations performed  
with Ising anisotropies typically reveal
no important differences with the results obtained with fully vector
models on finite systems. Small couplings in
the third direction play a similar role. As a consequence, 
for all practical purposes the critical behavior observed in the present studies describes
properly the expected physics of manganite models, which are always embedded in three
dimensional environments, and that have small anisotropies. 
A final note on this subject: The CE phase of manganites can show a finite-temperature
transition even in two dimensions, since the order parameter for charge order can be Ising type.

The CPU time needed to obtain the results shown in this subsection follows the
expected trends reported in previous investigations (see Table~\ref{tab:comp2D}). In particular,
the TPEM time needed for a 32$\times$32 cluster is comparable to the DIAG time on a 12$\times$12 cluster, 
a very encouraging result. Of course, this comparison will be even more favorable to the TPEM
with increasing number of CPU's for parallelization.

\begin{table}
\caption{\label{tab:comp2D} Comparison of the CPU times for the algorithms indicated, using an
Intel Pentium 4 (clock speed 3.06Gz) computer. Shown are results for different square
lattices of size $L\times L$,
assuming 2000 MC steps for thermalization,
and 2000 MC steps for measurements (taken every 10 MC steps). Since the TPEM can
be parallelized, some results were obtained using more than one CPU, as indicated.}
\begin{ruledtabular}
\begin{tabular}{ccccc}
$L$&Algorithm&\# of CPUs&CpuTime(h)\\
\hline
$12$&DIAG&$1$&$19.48$\\
$12$&TPEM&$2$&$5.46$\\
$20$&TPEM&$2$&$18.08$\\
$32$&TPEM&$8$&$25.92$\\
\end{tabular}
\end{ruledtabular}
\end{table}


\begin{figure}
\centerline{
\includegraphics[clip,width=9cm]{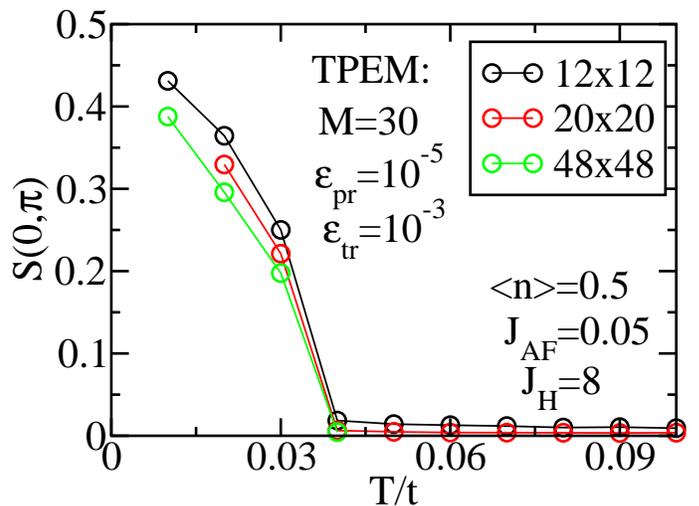}}
\caption{(color online). Spin structure factor vs. $T$ for
$J_{\rm AF}$=$0.05$ using the lattice sizes and parameters shown in the
figure.}
\label{Figure48}
\end{figure}

\subsection{Phase Diagram}

\begin{figure}
\centerline{
\includegraphics[clip,width=8cm]{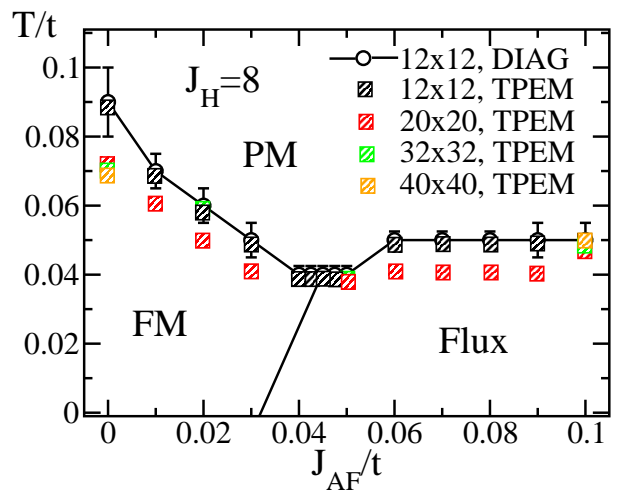}}
\caption{(color online). Phase diagram at $\langle n \rangle$=0.5, 
varying temperature and $J_{\rm AF}$.
Results are shown for a $12\times12$ lattice using both DIAG and TPEM techniques,
and for larger lattices using TPEM, as indicated. The origin of the tilting of the first-order
low-temperature FM-Flux line is explained in the text.} 
\label{Figure4}
\end{figure}

Using the TPEM, the phase diagram of the one-band model for manganites at $\langle n \rangle$=0.5 
was obtained (see Fig.~\ref{Figure4}). The transition between the FM and Flux states at low temperatures is of
first order. In fact, in the absence of quenched disorder the zero temperature result can be obtained by
using the perfect classical spin configurations for both the FM and Flux states, and calculating their
energy vs $J_{\rm AF}$ (not shown). By this procedure the zero-temperature 
critical $J_{\rm AF}$ was found to be close to 0.03. Raising the temperature, this transition line
is not vertical, but has a tilting.
Figure \ref{Figure4} shows that the estimated critical temperatures do not present
severe size effects, and the TPEM 
can be comfortably used at least up to 40$\times$40 clusters. The presence
of a first-order transition in the competition between the FM and Flux phases is in qualitative agreement
with several previous investigations that have shown similar trends both for the one and two bands models,
at any electronic density.\cite{review} This transition is expected to be severely affected by the influence of quenched
disorder, and this issue will be investigated in the near future. 

\begin{figure}
\centerline{
\includegraphics[clip,width=9cm]{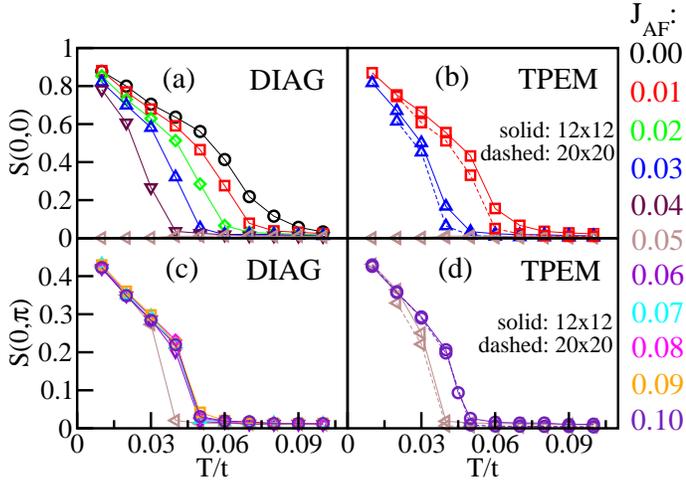}}
\caption{(color online). Examples of the criteria used in
the  calculation of the critical temperatures in Fig.~\ref{Figure4}. 
(a) and (c): Spin structure factors at the momenta of relevance vs. temperatures,
for many values of $J_{\rm AF}$, as indicated, using the DIAG technique on a $12\times12$ cluster.
(b) and (d): Same as in (a) and (c) except the technique used here is the TPEM. Shown are some of the results for
$12\times12$ and $20\times20$, as indicated in the figure.}
\label{Figure5}
\end{figure}

The critical temperatures in Fig.~\ref{Figure4} were estimated from the behavior of the spin structure factors
at the two momenta of relevance for the FM and Flux phases, as shown in Fig.~\ref{Figure5}.

\subsection{Density of States}

In this section, it is shown that the density-of-states (DOS) can be reproduced properly
by the TPEM. This is nontrivial, since it may be suspected that a method
based on an expansion of the DOS may have problems in an insulating phase due to the
rapid changes in the DOS near the gap. To our knowledge, this is the first time that
the TPEM is applied to an insulator. The discussion in this subsection shows
that the technique works satisfactorily.

\begin{figure}
\centerline{
\includegraphics[clip,width=7cm]{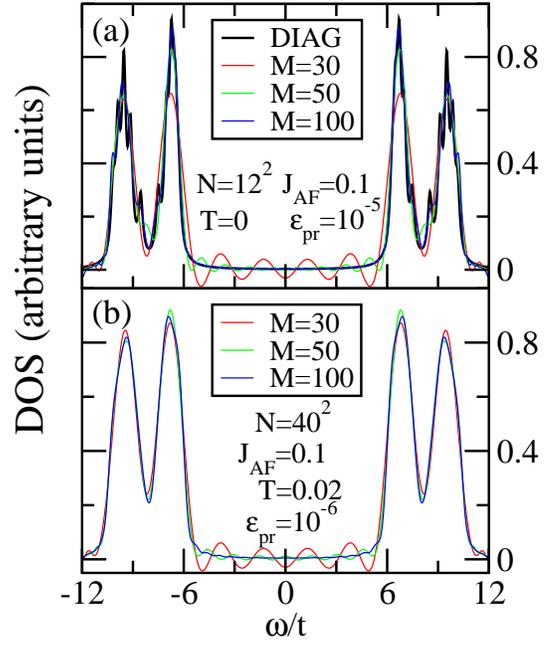}}
\caption{(color online). (a) Density-of-states calculated for a perfect Flux
state using both the DIAG and TPEM methods for a lattice of size $12\times12$.
In order to get the DOS accurately, even removing the Gibbs oscillations, 
one needs larger number of moments than what
is usually required for other observables. (b) Monte Carlo results for the 
density-of-states obtained from simulations performed on a 40$\times$40 lattice. In this
case the last configuration of the MC run has been used to calculate the
density-of-states at $T$=$0.02$.}
\label{dosfigure}
\end{figure}

In general, the density-of-states for a configuration of classical fields $\phi$ is given
by
\begin{equation}
N_{\phi}(\omega)=\sum_{\lambda}\delta(\omega^{\prime}-\epsilon_{\lambda}),
\end{equation}
with $\omega^{\prime}=(\omega-b)/a$, and 
where $a$ and $b$ are the parameters that normalize the
Hamiltonian in such a way that the new eigenvalues are in the interval $[-1,1]$.
These constants are given by
$a$=$(E_{\rm max} - E_{\rm min})/2$ and
$b$=$(E_{\rm max} + E_{\rm min})/2$,
where $E_{\rm max}$ and $E_{\rm min}$ are the maximum and minimum eigenvalues of the
Hamiltonian. Then, following the discussion of Section~\ref{sec:tpem},
the corresponding function $F(x)$ for the density-of-states
in the expression 
\begin{equation}
A(\phi)=\int_{-1}^{1}F(x)D(\phi,x)dx
\end{equation}
is the $\delta$-function
\begin{equation}
F(x)=\delta(\omega^{\prime}-x).
\end{equation}
In the expansion $F(x)$=$\sum_{m=0}^{\infty}f_{m}T_{m}(x)$, by using the expression
for the coefficients $f_m$=$\int_{-1}^{1}\alpha_{m}F(x)T(x)/(\pi\sqrt{1-x^{2}})$,
the final result for the density-of-states becomes
\begin{equation}
N_{\phi}(\omega)=\frac{\sum_{m}\alpha_{m}T_{m}(\omega^{\prime})\mu_{m}(\phi)}{\pi\sqrt{1-\omega^{'2}}}.
\end{equation}
This sum is truncated to a certain cutoff $M$ and such an abrupt truncation results
in unwanted Gibbs oscillations, as shown, e.g., in Fig.~\ref{dosfigure} for $M$=30.
This problem may be avoided by multiplying
the moments by dumping factors if needed.\cite{gonzalo-thomas}

The results for the density-of-states of 
the Flux phase are shown in Fig.~\ref{dosfigure}. 
Clearly, even at $M$=30 there is a very good agreement between the DOS calculated exactly
and with TPEM (with the exception of the in-gap Gibbs oscillations). Increasing $M$ further, 
even this effect disappears. Using a 40$\times$40 lattice, the results are almost
the same as those observed on the smaller system. It is concluded that the TPEM can
produce the DOS of insulating states accurately, and the method can be used to study
phase competition between metals and insulators.

\subsection{Conductances: Comparison TPEM vs. DIAG, and Results with Increasing Lattice Sizes}

\begin{figure}
\centerline{
\includegraphics[clip,width=9cm]{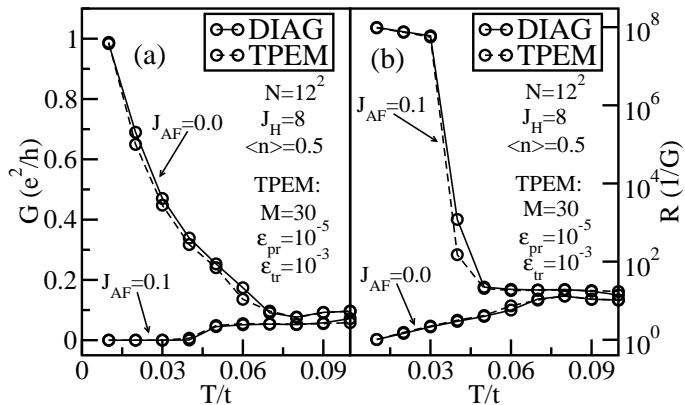}}
\caption{(a) Conductance and (b) resistance (1/conductance) vs. 
temperature for a $12\times12$ lattice, calculated with both 
DIAG and TPEM algorithms showing that the results agree.
The couplings used are $J_{\rm AF}$=$0.0$ and $J_{\rm AF}$=$0.1$ as indicated, and the density
is $\langle n \rangle$=0.5. The convergence at 
$T$=0.04 is achieved by using 40
moments with $\varepsilon_{\mbox{\scriptsize{{pr}}}}$=$10^{-7}$ 
and $\varepsilon_{\mbox{\scriptsize{{tr}}}}$=$10^{-5}$, as discussed in previous figures captions.} 
\label{Figure6}
\end{figure}

To compare theory with experiments, it is crucial to evaluate the conductance of the cluster under study.
Its temperature and magnetic field dependence will clarify whether the double-exchange models for
manganites contain the essence of the CMR phenomenon. The conductance calculation here follows the steps
previously extensively discussed by Verg\'es {\it et al.}, and it basically relies on the Landauer
formalism that links conductance with transmission. We refer the readers to original references
for more details (see for instance Ref.~\onlinecite{verges}). 
In Fig.~\ref{Figure6}, the conductance and its inverse (resistance) 
are shown as a function of temperature for the model on a 12$\times$12 cluster that can be solved both
with DIAG and TPEM. The agreement between the results obtained with both techniques is excellent,
at the two values of $J_{\rm AF}$ shown (one in the FM and the other in the Flux phase). Thus, 
the conductance calculation does not present an obstacle 
in the use of the TPEM .

\begin{figure}
\centerline{
\includegraphics[clip,width=9cm]{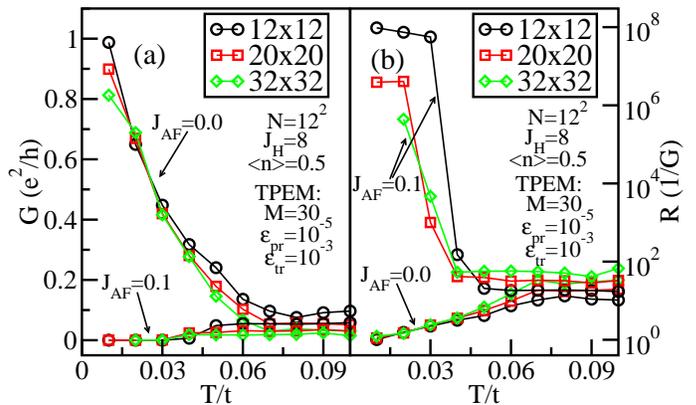}}
\caption{(color online). (a) Conductance and (b) resistance (1/conductance) 
vs. temperature calculated with the 
TPEM algorithm at $J_{\rm AF}=0.0$ and $J_{\rm AF}=0.1$, 
$\langle n \rangle$=0.5, and
for the cluster sizes shown in
the figure. The figure shows that the conductance does $not$ suffer from
strong size effects.
The convergence at $T$=0.04 for the $12\times12$ lattice was achieved by using 40
moments with $\varepsilon_{\mbox{\scriptsize{{pr}}}}=10^{-7}$ and $\varepsilon_{\mbox{\scriptsize{{tr}}}}=10^{-5}$, as discussed elsewhere.} 
\label{Figure7}
\end{figure}

With increasing lattice size, the TPEM conductance behaves smoothly and the finite-size
effects are small (see Fig.~\ref{Figure7}), with the only exception of the insulating Flux
phase regime at low temperatures where the 12$\times$12 cluster results appear appreciably
different from those on larger systems. Considering the small value of the conductance 
in this insulating regime and the subsequent convergence of the Flux-phase resistance 
between the 20$\times$20 and 32$\times$32 clusters, this appears to be only a minor issue.

\subsection{Influence of Magnetic Fields in the Clean Limit and 
Partial Conclusions}

\begin{figure}
\centerline{
\includegraphics[clip,width=7cm]{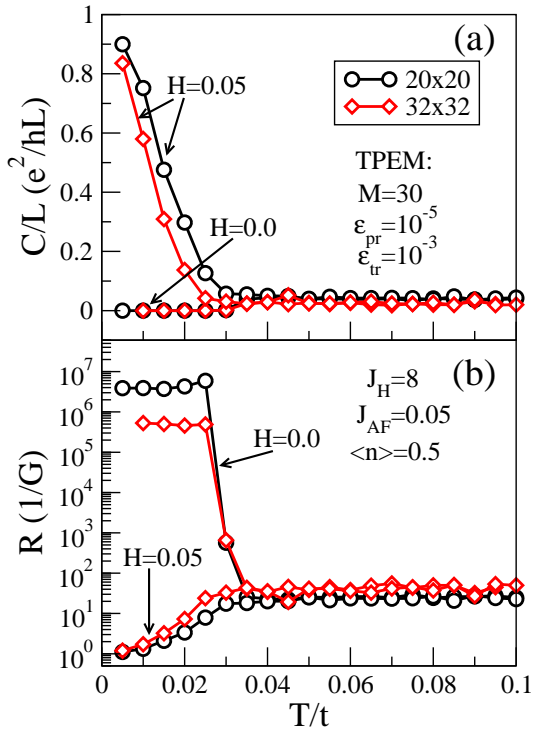}}
\caption{(color online). (a) Conductance vs. temperature for lattices of sizes
$20\times20$ and $32\times32$, with and without magnetic fields. (b) Resistance
vs. temperature calculated by taking the inverse of the conductance in (a).
Close to the first-order transition in the phase diagram, a small magnetic field 
can destabilize the insulating Flux state into a metallic FM state. 
For each temperature, 1000 thermalization and 2000 measurements MC
steps were used, with actual measurements 
taken at every 10 steps.}
\label{resmagneticfigure}
\end{figure}

As discussed in the introduction, it is important to investigate if the
models studied here, in the clean limit, present a large magnetoresistance effect.
Previous studies by 
Aliaga {\it et al.} \cite{aliaga} on 4$\times$4 clusters, suggested 
that the ``low temperature'' large magnetoresistance
experimentally observed in some manganites \cite{low-T-CMR} 
can be explained by double-exchange models in the clean limit. This result
is important and deserves to be confirmed using larger clusters. Here, the
case of the one-band model is analyzed, with results shown in Fig.~\ref{resmagneticfigure}
(two bands will be studied later in this paper). The value of $J_{\rm AF}$ 
was chosen to be on the insulating side (Flux phase) of the phase diagram Fig.~\ref{Figure4},
but close to the first-order transition separating the metal from the insulator.
The application of `small' magnetic fields favors the FM state over the Flux state
and that manifests as a sharp transition from the metal to the insulator, for
values of the magnetic field that appear abnormally small in the natural units
of the problem. Thus, this model presents a huge negative magnetoresistance,
an encouraging result that shows theory is in the right track to understand manganites.
The effect shown in Fig.~\ref{resmagneticfigure} is caused by the proximity in energy of two states
with quite different properties, i.e. there is a {\it hidden small energy scale} in the
problem.

However, note that
the  standard large finite-temperature magnetoresistance traditionally studied in
Mn-oxides cannot be understood with clean limit models, as shown 
in Fig.~\ref{resmagneticfigure}: the zero magnetic-field resistivity does not
have the large peak near Curie temperatures characteristic of CMR manganites. Future work using 
the TPEM will analyze whether this more traditional CMR effect can be obtained
including quenched disorder.

\vskip 0.4cm

Overall, it can be safely concluded that the study of the one-band model with the TPEM
has proven that the technique works properly, and that the FM vs. Flux competition 
occurs via a first-order metal-insulator transition. This regime is ideal for the
analysis of the influence of quenched disorder in future calculations.

\section{RESULTS FOR THE TWO-BAND MODEL}\label{sec:twoband}

\subsection{Definition}

In this effort, the two-band model for manganites was also investigated using the TPEM.
The two bands arise from the $e_{\rm g}$ bands that are active at the Mn ions in Mn-oxides,
as extensively discussed before.\cite{review} The overall conclusion of this section is that the TPEM
is also a good approximation to carry out computational studies, conclusion similar to that
reached for only  one active band.
The Hamiltonian for this model is\cite{review}

\begin{eqnarray}
H_{2b}&=&\sum_{\gamma,\gamma',i,\alpha}t^\alpha_{\gamma\gamma'}
{\mathcal S}(\theta_i,\phi_i,\theta_{i+\alpha},\phi_{i+\alpha})
c^\dagger_{i,\gamma}c_{i+\alpha,\gamma'} \nonumber \\ &+& 
\lambda\sum_{i}(Q_{1i} \rho_i +  Q_{2i} \tau_{xi} + Q_{3i} \tau_{zi}) \nonumber \\&+&
\sum_{i}\sum_{\alpha=1}^{\alpha=3} D_\alpha Q_{\alpha i}^2,
\label{eq:hamtwobands}
\end{eqnarray}
where the factor that renormalizes the hopping in the $J_{\rm H}$=$\infty$ limit is
\begin{eqnarray}
{\mathcal S}(\theta_i,\phi_i,\theta_{j},\phi_{j})&=&\cos(\frac{\theta_{i}}{2})\cos(\frac{\theta_{j}}{2})
\nonumber \\&+&\sin(\frac{\theta_{i}}{2})\sin(\frac{\theta_{j}}{2})e^{-i (\theta_{i}-\theta_{j})}.
\end{eqnarray}
The parameters $t^\alpha_{\gamma\gamma'}$ are the 
hopping amplitudes 
between the orbitals $\gamma$ and $\gamma'$ in the direction $\alpha$. In this paper, we restrict ourselves
to two dimensions, such that $t^{x}_{aa}=-\sqrt{3}t^{x}_{ab}=
-\sqrt{3}t^{x}_{ba}=3t^{x}_{bb}=1$,
and $t^{y}_{aa}=\sqrt{3}t^{y}_{ab}=\sqrt{3}t^{y}_{ba}
=3t^{y}_{bb}=1$. $Q_{1i}$, $Q_{2i}$ and $Q_{3i}$ are normal 
modes of vibration that can be expressed in terms of the oxygen coordinate $u_{i,\alpha}$ as:
\begin{eqnarray}
Q_{1i}&=&\frac {1}{\sqrt{3}} [(u_{i,z}-u_{i-z,z}) + (u_{i,x}-u_{i-x,x}) \nonumber \\ &+&  
(u_{i,y}- u_{i-y,y})], \nonumber \\
Q_{2i}&=&\frac {1}{\sqrt{2}} (u_{i,x}-u_{i-x,x}), \nonumber \\
Q_{3i}&=&\frac {2}{\sqrt{6}} (u_{i,z}-u_{i-z,z}) 
 -\frac{1}{\sqrt{6}} (u_{i,x}-u_{i-x,x}) \nonumber \\&-&\frac{1}{\sqrt{6}} (u_{i,y}- u_{i-y,y}). \nonumber
\end{eqnarray}
Also, $\tau_{xi}=c^{\dagger}_{ia}c_{ib}+c^{\dagger}_{ib}c_{ia}$, 
$\tau_{zi}=c^{\dagger}_{ia}c_{ia}-c^{\dagger}_{ib}c_{ib}$, and 
$\rho_{i}=c^{\dagger}_{ia}c_{ia}+c^{\dagger}_{ib}c_{ib}$.
The constant $\lambda$ is the electron-phonon coupling
related to the Jahn-Teller distortion 
of the MnO$_6$ octahedron.\cite{tokura,review} 
Regarding the phononic stiffness, and in
units of $t^{x}_{aa}=1$, the $D_{\alpha}$ parameters are $D_1=1$ and $D_2=D_3=0.5$,
as discussed in previous literature.\cite{aliaga} 
The rest of the notation is standard. In our effort here, the emphasis is on the case $\lambda$=0
believed to be of sufficient relevance to deserve a special study since it already contains\cite{aliaga} a
competition between FM metallic and CE insulating states at $\langle n \rangle$=0.5. Thus,
this is an excellent testing ground for the TPEM, particularly having in mind the next
challenge involving a TPEM study in the presence of quenched disorder.
However, briefly some results at nonzero $\lambda$ will also be shown.


\subsection{TPEM Performance}

\subsubsection{Test of the TPEM in Small Systems}

\begin{figure}
\centerline{
\includegraphics[clip,width=9cm]{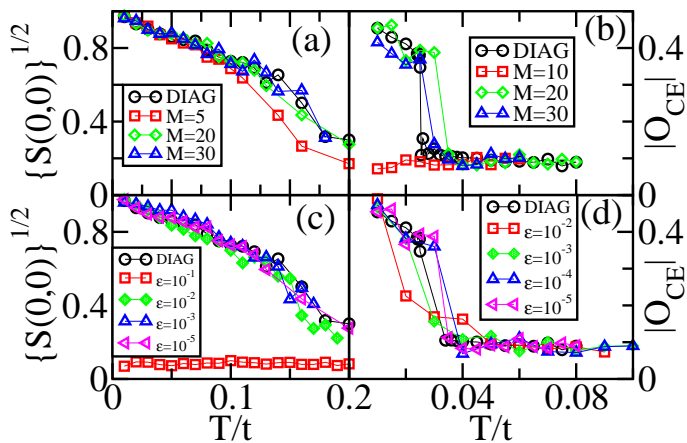}}
\caption{(color online). Convergence of the structure factors, varying the parameters of the TPEM algorithm: 
(a) Magnetization $|M|$=$\sqrt{S(0,0)}$ vs. $T/t$ at $J_{\rm AF}$=$0.0$ using the DIAG method 
and TPEM with $\epsilon_{\rm pr}$=$10^{-5}$, $\epsilon_{\rm tr}$=$10^{-6}$, and the values of 
$M$ indicated. (b) Order parameter associated with the CE phase $|O_{\rm CE}|$=$\sqrt{S(\pi,0)}$ 
vs. $T/t$ at $J_{\rm AF}$=$0.2$ using the DIAG method and TPEM with $\epsilon_{\rm pr}$=$10^{-5}$, 
$\epsilon_{\rm tr}$=$10^{-6}$, and the values of $M$ indicated. (c) Magnetization 
$|M|$=$\sqrt{S(0,0)}$ vs. $T/t$ at $J_{\rm AF}= 0.0$ using the DIAG method and TPEM with 
$M$=$20$ and $\epsilon_{\rm tr}=10^{-6}$, varying $\epsilon_{\rm pr}$=$\epsilon$ as indicated. 
(d) Order parameter of the CE phase $|O_{\rm CE}|$=$\sqrt{S(\pi,0)}$ vs. $T/t$ at $J_{\rm AF}= 0.2$ 
using the DIAG method and TPEM with $M$=$20$, $\epsilon_{\rm tr}$=$10^{-6}$, varying 
$\epsilon_{\rm pr}$=$\epsilon$ as indicated. All calculations were done on a $12\times12$ 
lattice, using $1000$ Monte Carlo steps for thermalization and $1000$ for measurements.}
\label{Figure8}
\end{figure}

As in the case of the one-band model, the analysis starts here by comparing DIAG and TPEM 
results on small systems. Figure \ref{Figure8} contains the magnetization (in absolute value,
and coming from the classical spins) vs. temperature. The results in (a) and (c) were obtained
at $J_{\rm AF}$=0.0, $\lambda$=0, and $\langle n \rangle$=0.5, a regime known to develop
ferromagnetism at low temperatures.\cite{aliaga} Indeed, both methods show a nonzero value for the
magnetization. The dependence with the TPEM parameters indicates that a $M$ of approximately
30 or higher is sufficient to get accurate results. This is a conclusion that also appears
in (b) where the case of a CE state is studied, which is stabilized with increasing $J_{\rm AF}$.
Regarding the other TPEM parameters, (d) shows that $\epsilon_{\rm pr}$=10$^{-5}$, as used for the
one-band case, leads to accurate results. Overall, it seems that the same set of
parameters deduced from the one-band model investigations can also be used for two bands,
an interesting simplifying result.

\subsubsection{Dependence of Results on Lattice Sizes}

\begin{figure}
\centerline{
\includegraphics[clip,width=9cm]{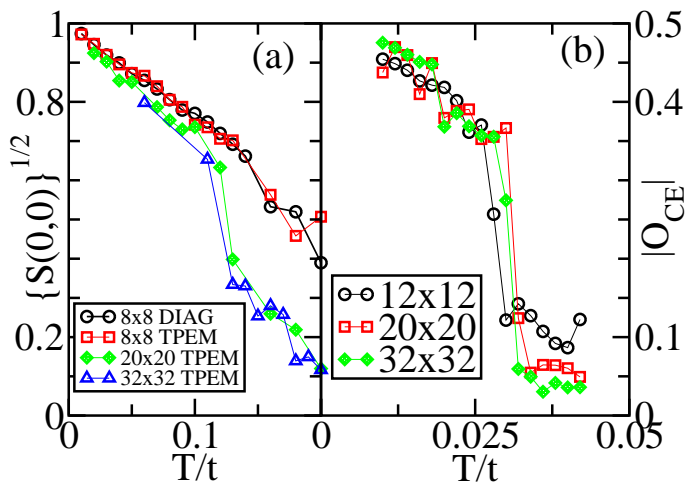}}
\caption{(color online). Lattice size 
dependence of the square root of the structure 
factors at the momenta characteristic of (a) a FM state, ${\mathbf k}$=$(0,0)$ 
and $J_{\rm AF}$=$0.0$, and (b) a CE phase, ${\mathbf k}$=$(\pi,0)$ and 
$J_{\rm AF}$=$0.2$. Results were obtained with the TPEM with $M$=$20$, 
$\epsilon_{\rm pr}$=$10^{-5}$, $\epsilon_{\rm tr}$=$10^{-6}$. In addition, 
for (a) the DIAG method was also used on a $8\times8$ lattice as indicated. 
In the simulation, $1000$ MC steps were used for thermalizations and $1000$ 
steps for measurements. }
\label{Figure9}
\end{figure}

Figure \ref{Figure9} illustrates the dependence of results on lattice sizes. The systematic
behavior is similar to that observed in the case of the one-band model at the same density $\langle n \rangle$=0.5.
In (a) results for the magnetization vs. temperature indicate the existence of a FM state at
low temperatures, as well as small finite-size effects when the 20$\times$20 and 32$\times$32 clusters
are compared. Even less pronounced size effects are found in the CE regime, increasing $J_{\rm AF}$
as shown in (b). There is no indication that the TPEM deteriorates with increasing lattice
size, providing hope that this method will be strong enough to handle the introduction of quenched
disorder in future studies.

\subsection{Phase Diagram}

\subsubsection{Results without Phonons}

\begin{figure}
\centerline{
\includegraphics[clip,width=9cm]{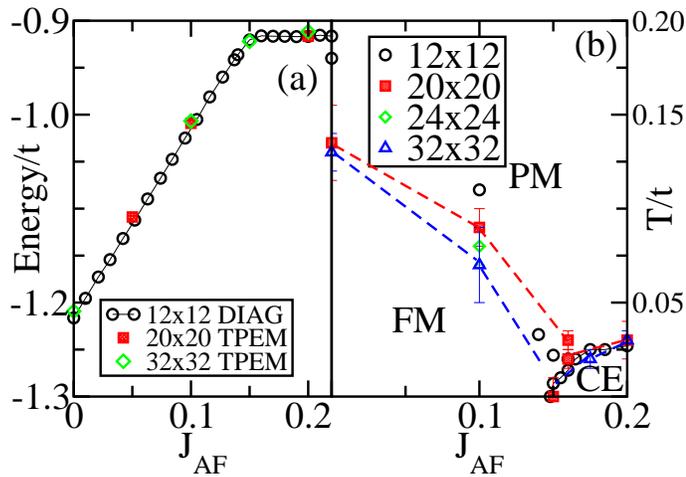}}
\caption{(color online). (a) Total Energy vs. $J_{\rm AF}$ at low temperature ($T$=$0.01t$) for different lattices 
as indicated. For the $12\times12$ lattice the DIAG method was used, while for the others the TPEM 
was employed with $M$=$20$, $\epsilon_{\rm pr}$=$10^{-5}$, $\epsilon_{\rm tr}$=$10^{-6}$. 
In the simulation, $1000$ MC steps were used for thermalizations and $1000$ steps for measurements. 
(b) Phase diagram of Hamiltonian Eq.~(\ref{eq:hamtwobands}) varying temperature and $J_{\rm AF}$ ($\lambda$=0). 
The three magnetically different regions: FM, PM, and CE are indicated. The phase diagram was 
calculated for different lattices as shown. For $12\times12$ the DIAG method was used and for the 
others the TPEM with $M$=$20$, $\epsilon_{\rm pr}$=$10^{-5}$, $\epsilon_{\rm tr}$=$10^{-6}$. The 
critical temperatures were obtained from the calculation of structure factors, as shown in 
Fig.~\ref{Figure9}.} 
\label{Figure10}
\end{figure}

To further test the TPEM, the phase diagram of the two-band model at $\lambda$=0
and $\langle n \rangle$=0.5 was obtained. Also at very low temperature, the energy
was found as a function of $J_{\rm AF}$. The results are in Fig.~\ref{Figure10}.
Part (a) shows an excellent agreement among the several lattice sizes studied here.
The abrupt change in the slope of the curve near $J_{\rm AF}$=0.15 indicates a
first-order transition, similar to that found in the one-band case and in 
previous literature. In (b), the full phase diagram is obtained. There are clear
qualitative similarities with the results presented before by Aliaga {\it et al.}
using a 4$\times$4 cluster.\cite{aliaga} In particular, 
the curve defining the CE phase at low temperature
has a positive slope rather than being vertical as in other cases. The fact that
the TPEM gives results in excellent agreement with DIAG but on substantially
larger systems is very encouraging and establishes this technique as a key method
for a frontal attack to the CMR problem using realistic models and quenched disorder.

For completeness, CPU times for the case of the two-band model are provided
in Table \ref{Table2}. The CPU time per site does not change dramatically
with $N$, close to the expected theoretical estimation for the TPEM. Clearly, lattices
well beyond 32$\times$32 can be handled with this technique.

\begin{table}[ht]
\caption{\label{Table2} CPU Times of the TPEM in seconds per 5 Monte Carlo steps for 
Hamiltonian Eq.~(\ref{eq:hamtwobands}) with $J_{\rm AF}$=$0$ and inverse
temperature $\beta=50$, and the lattices shown. 
The third column is the ratio of CPU time per lattice site. The computer used was an AMD 
Opteron(tm) 244, 1.8GHz with 1MB cache. The TPEM parameters were
$M$=20, $\epsilon_{\rm pr}$=$10^{-5}$, and
$\epsilon_{\rm tr}$=$10^{-6}$.}
\begin{ruledtabular}
\begin{tabular}{ccc}
$L\times L$&CPU Time(s)&CPU Time/N\\
\hline
$12 \times 12$&124&$0.86$\\
$20 \times 20$&642&$1.61$\\
$24 \times 24$&992&$1.72$\\
$32 \times 32$&2451&$2.39$\\
\end{tabular}
\end{ruledtabular}
\end{table}


\subsubsection{Influence of Phonons}

\begin{figure}
\centerline{
\includegraphics[clip,width=9cm]{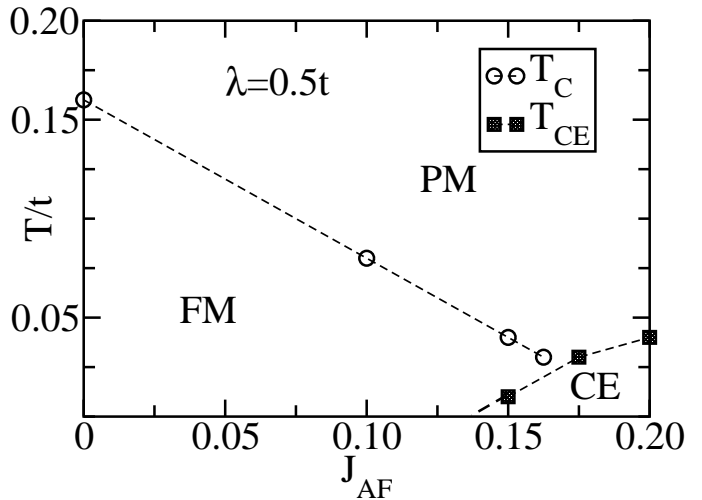}}
\caption{Phase diagram of Hamiltonian Eq.~(\ref{eq:hamtwobands}) varying temperature and $J_{\rm AF}$ 
for $\lambda$=$0.5$ and the harmonic parameters discussed in the text, 
i.e., with the inclusion of phonons. The 
critical temperatures were obtained from the calculation of structure factors on a $12\times12$ lattice 
with the DIAG method. Note the similarity of this phase diagram with the result obtained at 
$\lambda$= 0.0, suggesting that to simulate the competition between FM metallic and CE insulating 
regimes the presence of a robust electron-phonon coupling is not necessary.} 
\label{Figure11}
\end{figure}

As explained in the introduction, the general scenario proposed for manganites
does not depend on particular details of the competing phases, but in the
competition itself.\cite{review} Thus, to the extent that the FM metallic and CE insulating
phases are found in competition, the value of electron-phonon coupling
 $\lambda$ is not of crucial relevance. We believe that $\lambda$ is likely small
in practice, since recent experiments are not finding evidence of a robust
charge checkerboard (see, for example, Ref.~\onlinecite{noCO})
and, in addition, theoretical studies have shown that a large
$\lambda$ renders the FM state also insulating.\cite{aliaga} 
Nevertheless, to confirm that the
results are not severely affected by switching on $\lambda$, in Fig.~\ref{Figure11}
the phase diagram for $\lambda$=0.5 is presented on a lattice substantially larger
than used in previous investigations.\cite{aliaga} Comparing Figs.~\ref{Figure10} and \ref{Figure11},
clearly both cases lead to very similar phase diagrams. Since removing the phononic
degree of freedom speeds up the simulations, these results suggest that the future
effort in this context could focus in the $\lambda$=0 case and still expect to
find realistic conclusions.

\subsection{Density of States}

\begin{figure}
\centerline{
\includegraphics[clip,width=9cm]{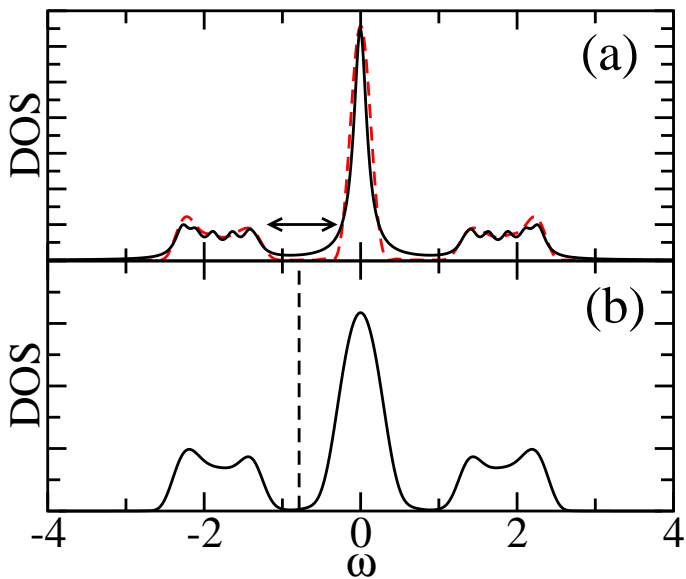}}
\caption{
(color online) (a) DOS of a perfect CE phase (single spin configuration) 
obtained on a 12$\times$12 lattice
calculated with the DIAG method (solid black line) 
and with the TPEM (dashed red
line) using $M$=$100$. The chemical potential lies in the left gap (arrows)
indicating that the system is an insulator.
(b) DOS of the system with $J_{\rm AF}$=$0.2$ 
(CE-phase ground state) on a 20$\times$20
lattice calculated with the TPEM using $M$=$100$, as
described in the text. The location of the chemical potential is indicated by
the vertical dashed line.
} 
\label{FigureDOS}
\end{figure}

As in the case of the one-band model, we also tested whether the TPEM technique can
reproduce the DOS of the two-band model 
in the regime where the system is insulating (CE phase). The results are in
Fig.~\ref{FigureDOS}. Part (a) shows a comparison between DIAG and TPEM on
a 12$\times$12 cluster. The agreement is excellent for the case of $M$=100 (shown),
and fairly acceptable for smaller values of $M$. For larger lattices that
can only be studied with TPEM (part (b)), the results are also in good
agreement with
expectations.
Then, no problems have been detected in calculating the DOS using the TPEM technique
in the regime where the model is in an insulating state.

\subsection{Conductances: Comparison TPEM vs. DIAG, 
and Results for Increasing Lattice Sizes}

\begin{figure}
\centerline{
\includegraphics[clip,width=9cm]{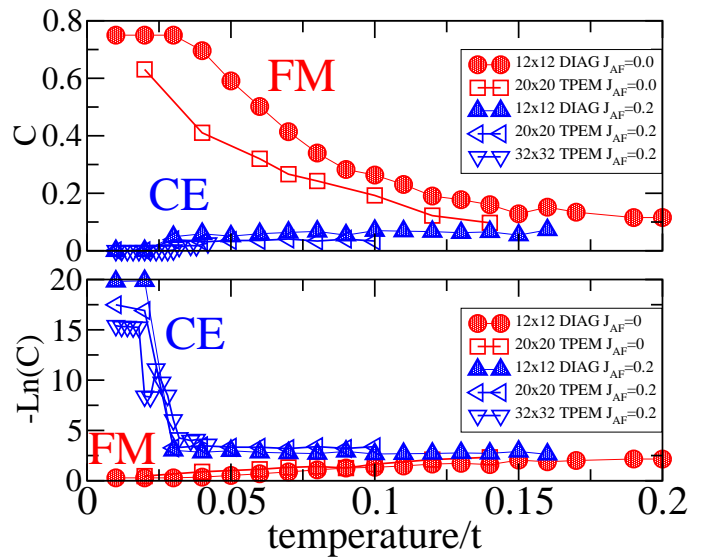}}
\caption{(color online) {\emph {Upper Panel:} } Conductance vs.
temperature for $J_{\rm {AF}}=0.0$
(ferromagnet) and $J_{\rm {AF}}=0.2$ (CE phase), different lattices
sizes and algorithms, as indicated.
{\emph {Lower Panel:}} Logarithm of the resistivity vs. temperature
for the same parameters as
before.}
\label{Figure12}
\end{figure}

As in the case of the one-band model, the final test for the two-band case 
is the calculation of the conductance. Results are shown in Fig.~\ref{Figure12}.
Even using clusters much larger than can be handled with DIAG, the behavior
of the conductance is properly captured by TPEM. There are no hidden subtleties involved
in this estimation of transport properties, opening the way toward many future
applications of this technique.

\subsection{Influence of Magnetic Fields in the Clean Limit}

\begin{figure}
\centerline{
\includegraphics[clip,width=9cm]{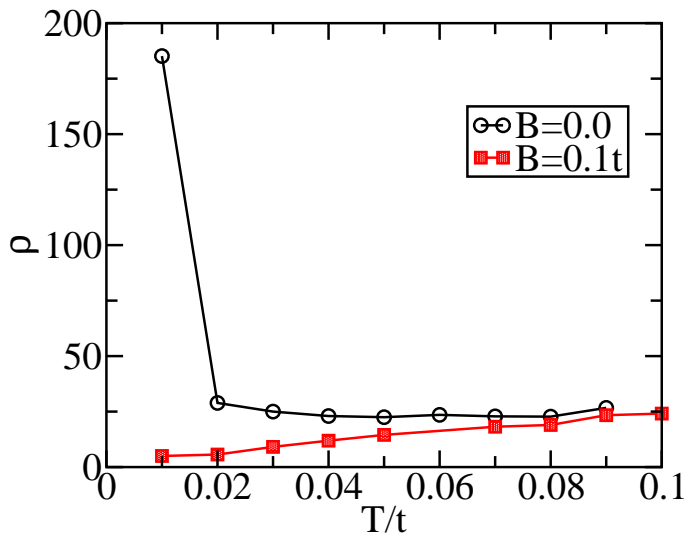}}
\caption{(color online) Effect of a small magnetic field on the resistivity 
of the CE phase, for couplings in the vicinity of the first order
transition metal-insulator. Shown is the
resistivity vs. $T$ at $J_{\rm AF}$=$0.175$, $\lambda$=0,  
on a 20$\times$20 lattice for
$B$=$0.1t$ and without field ($B$=$0$) for comparison. Note the large
change in the resistivity at low temperature, compatible with a colossal magnetoresistance.
The TPEM was used with
$M$=$30$, $\epsilon_{\rm pr}$=$10^{-5}$ and $\epsilon_{\rm tr}$=$10^{-6}$. 
1,000 Monte Carlo steps
were done for thermalization and an additional 100 more for measurements.
}
\label{FigureMAG}
\end{figure}

Similarly as for the case of the one-band model, we have also
studied the influence of a magnetic field in the case of the two-band
model. The value of $J_{\rm AF}$ was chosen to be 0.175, namely on the
CE side but close to the first-order metal-insulator transition. As shown
in Fig.~\ref{FigureMAG}, the application of a relatively small field
-- in the natural units of the problem -- leads to a drastic change
in the resistivity at low temperatures. In this regime, the insulator is
transformed into a metal (negative
magnetoresistance). As remarked before,
this is in agreement with previous studies carried
out by Aliaga {\it et al.} \cite{aliaga}, showing
that the ``low temperature'' large magnetoresistance materials\cite{low-T-CMR}
can be understood by double-exchange models in the clean limit. However, as
in the case of one-band models studied before,
the finite-temperature CMR effect in
Mn-oxides cannot be understood with clean limit models, as shown 
in Fig.~\ref{FigureMAG}, since the zero magnetic-field resistivity does not
have the large peak characteristic of manganites. Future work with TPEM
will address this issue including quenched disorder.

\section{CONCLUSIONS}

In this paper, it has been shown that the computationally 
intensive exact-diagonalization algorithm  for the study 
of CMR-manganite models can in practice 
be replaced by the novel Truncated Polynomial Expansion Method (TPEM)
without any substantial loss in accuracy.
The DIAG, although being exact, does not permit simulations 
of large clusters owing to the fact that the computational cost
grows like the $4^{th}$ power of the system size, $N$. 
On the other hand, the newly developed TPEM algorithm reduces the computational
complexity to $\mathcal{O}(N)$, thereby allowing for simulations of 
larger systems. 

For both the one- and two-band 
double exchange model with interacting Mn spins, we have compared systematically the results calculated with both the diagonalization method and the TPEM algorithm. 
As the spin-spin coupling $J_{\rm AF}$ is varied, 
the one-band model reveals a low-temperature first-order
phase transition between conducting FM and 
insulating Flux states 
in the vicinity of $J_{\rm AF}$=$0.045$. For two bands,
a similar first-order transition separates FM metallic and CE insulating phases. 
Our calculations presented here 
included a systematic study of the performance of the TPEM algorithm varying
its parameters $M$, $\epsilon_{\rm pr}$ and $\epsilon_{\rm tr}$, already defined
in the introductory sections. It was shown that the results of the TPEM 
algorithm converge to those of
the DIAG algorithm with increasing $M$, and for $M \ge 30$ 
TPEM results
are sufficiently accurate to obtain phase diagrams
with small error bars. This number appears to be fairly stable
under changes in the model, couplings, and for different phases,
and lattice sizes. Also nothing indicates that working at density different 
from 0.5, the focus of the current effort, will spoil the TPEM performance.
[However, it is advisable to
be particularly cautious near critical temperatures, where in some cases we
found the need to increase $M$ to 40]. 
Similar systematic results were presented
for the $\epsilon$'s. Overall, the general process of fixing TPEM parameters on small
systems by comparing with DIAG and then using the same parameters on larger
lattices appears reliable, and this method will be applied to other systems
in the near future.

Parallelization of the TPEM algorithm makes it possible 
to study large clusters. Taking advantage of this possibility, 
in the absence of quenched disorder, 
we have made calculations on lattices of up to $40\times40$ sites even for the case
of a finite Hund coupling in the one-band model. 
But previous calculations in the limit of $J_{\rm H}$=$\infty$ used up to 8000
sites,\cite{heisenberg} thus even accounting for the factors of 2 involved in
comparing finite and infinite $J_{\rm H}$ there is still room for further
improvement. We have also checked that calculations of conductances,
crucial to predict transport properties, also can be carried out smoothly
with the TPEM, and in addition the size effects are in general small.
While obviously the increase in lattice size allows for
more accurate determinations of critical temperatures, even more importantly these
large lattices will be crucial for the next big step in large-scale
manganite simulations which
is the introduction of quenched disorder. This disorder is expected to lead
to a percolative-like picture that causes the CMR phenomenon. Any percolative
effect requires large systems, and having access to clusters substantially larger than
those studied with DIAG is a key necessary condition to unveil the conceptual
reason behind the CMR phenomenon.

Interesting physical results were also here reported. This includes the 
one-band phase diagram with a metal-insulator transition. But the main result
in this context is the presence of an enormous magnetoresistance effect at
low temperatures, even in the clean limit studied here. This effect was already
anticipated by Aliaga {\it et al.} in their pioneering work on this subject on small
systems.\cite{aliaga}
The survival of the effect on the large clusters reachable by the 
TPEM, as described in this manuscript, shows that {\it some forms of
CMR found experimentally can already be accurately
reproduced using realistic models}, providing
further support that a theoretical solution of the CMR puzzle is within reach.
But for the most common form of CMR in Mn-oxides at temperatures close
to the Curie temperature, studies with quenched
disorder will likely be needed.

Summarizing, we here reported a successful implementation of the TPEM for the
study of double-exchange-like
models for manganites. The technique has a CPU time that grows linearly
with the number of sites, and in addition it is parallelizable. 
Thus, the main result is that a novel technique has
been identified and tested
that can led to a frontal attack of the most interesting problem
in manganites: the analysis of large magnetoresistance effects in the presence
of quenched disorder, when phases compete via a first-order transition in the
clean limit. 
This ``holy grail'' of simulations will be the focus of our effort in the
near future. It will demand at least an order of magnitude
more effort than in the present manuscript, but this can be alleviated by
increasing the number of nodes available for the simulations and we are already
working on this aspect. The large-scale computational 
facilities at Oak Ridge National Laboratory 
will play a key role in reaching this ambitious goal.

\section{ACKNOWLEDGMENTS}
Most of the computational work in this effort was performed 
at the supercomputing facilities of the Center for Computational Science at 
the Oak Ridge National Laboratory (ORNL), managed by UT-Battelle, LLC, for the 
U.S. Department of Energy under Contract DE-AC05-000R22725.
This work is supported by the LDRD
program at ORNL and by the NSF grant DMR-0443144.
We also acknowledge the help of J. A. Verg\'{e}s in the study of conductances. 
This research used the SPF computer program and software toolkit developed at ORNL (http://mri-fre.ornl.gov/spf).


\end{document}